\documentclass[runningheads,a4paper]{llncs}
\usepackage{amssymb} 
\usepackage{graphicx}
\usepackage{url}
\usepackage{float}
\usepackage{soul,color}
\usepackage{todonotes}

\usepackage{balance}		
\usepackage[utf8]{inputenc}
\usepackage{newunicodechar}
\newunicodechar{ﬁ}{fi}
\newunicodechar{ﬀ}{ff}
\usepackage[utf8]{inputenc}
\usepackage[T1]{fontenc}
\usepackage{booktabs}
\usepackage{xcolor}
\usepackage{url}
\usepackage{multirow}
\newcommand{\quotes}[1]{``#1''}
\usepackage{enumitem}
\usepackage{amssymb} 

\begin{document}

\title{Cyber-Storms Come from Clouds: Security of Cloud Computing in the IoT Era}

\authorrunning{Michele De Donno et al.}

\titlerunning{Cyber-Storms Come from Clouds}

\author{
Michele De Donno\inst{1},
Juxhino Kavaja \inst{1},
Nicola Dragoni \inst{1,2}, 
Antonio Bucchiarone \inst{3},
Manuel Mazzara \inst{4} \\
\institute{DTU Compute, Technical University of Denmark, Denmark \\
\{juxk, mido, ndra\}@dtu.dk\\
\and Centre for Applied Autonomous Sensor Systems, \"{O}rebro University, Sweden \\
nicola.dragoni@oru.se
\and Fondazione Bruno Kessler, Trento, Italy \\
bucchiarone@fbk.eu
\and Innopolis University, Russian Federation\\
m.mazzara@innopolis.ru}}

\maketitle

\abstract{The Internet of Things (IoT) is rapidly changing our society to a world where every “thing” is connected to the Internet,
making computing pervasive like never before. This tsunami of connectivity and data collection relies more and more on the Cloud,
where data analytics and intelligence actually reside. Cloud computing has indeed revolutionized the way computational resources and
services can be used and accessed, implementing the concept of utility computing whose advantages are undeniable for every
business. However, despite the beneﬁts in terms of flexibility, economic savings, and support of new services, its widespread adoption
is hindered by the security issues arising with its usage. From a security perspective, the technological revolution introduced by IoT and
Cloud computing can represent a disaster, as each object might become inherently remotely hackable and, as a consequence,
controllable by malicious actors. While the literature mostly focuses on security of IoT and Cloud computing as separate entities, in this
article we provide an up-to-date and well-structured survey of the security issues of Cloud computing  in the IoT era. We give a clear picture of where security issues occur and what their potential impact is. As a result, we claim that it is not enough to secure IoT devices, as cyber-storms come from Clouds.}


\section{Introduction}

The Internet of Things (IoT) is rapidly and inevitably spreading in our society, with the promise of rising efficiency and connectivity. Although the number of “things” has strongly been increasing over the past few years, statistics predict an even further growth in the future. Indeed, if the number of IoT connected devices in 2017 was around 20
billion, there will be about 30 billion in 2020 and more than double in 2025 \cite{IoTdevs}. This dramatic increase will bring challenges together with opportunities, and the massive introduction of this technology will need to be managed by several points of views such as legal, social, business-wise and of course technological \cite{Rehman2017}.

IoT applications span from industrial automation to home area networks to smart building, pervasive healthcare and smart transportation \cite{Nalin2016,Salikhov2016a,Salikhov2016b}. For instance, smart homes will heavily rely upon IoT devices to monitor the house temperature, possible gas leakages, malicious intrusions, and several other parameters concerning the house and its inhabitants. In pervasive healthcare, IoT devices are used to perform continuous biological monitoring, drug administration, elderly monitoring conditions and habits for improved lifestyle, and so on. Last but not least, with the Industry 4.0 technological revolution, Industrial IoT (IIoT) is entering its golden age.

From a security perspective, this plethora of IoT devices flooding the world is having tremendous consequences, so that it is not an exaggeration to talk about a security and privacy disaster \cite{DraGiaMaz_SEDA16}. In fact, IoT devices are often bad or not protected at all, thus, easily exploitable from different families of malwares to perpetrate large scale attacks (this is the case of Distributed Denial of Service-Capable IoT malwares such as Mirai \cite{mido_2018,DonnoDGM16}, just to mention a key example).

If we refer to one of the most common definitions of IoT, we can see that it is based on a single layer of devices with embedded computation and connectivity: \quotes{the interconnection via the Internet of computing devices embedded in everyday objects, enabling them to send and receive data} \cite{IoT_Definition}. This definition depicts the traditional scenario which most of the literature about IoT security focuses on (\cite{Yang_2017,Lin_2017,Conti_2018}, just to mention a few papers). Nevertheless, focusing only on the security of end devices risks to make us lose the sight of the overall picture.\\

\textit{There Is No IoT without the Cloud.} Today, IoT systems strongly rely on the Cloud. End devices are increasingly used as lightweight devices that collect data and connect to powerful Cloud servers responsible for all the application intelligence and data analytics \cite{Guan_2017,Mihovska_2018,Malik_2018}. This huge amount of data sent to the Cloud is one of the main motivations for the investigation of new distributed computing paradigms, such as Fog Computing \cite{Mahmud_2018}.

For this reason, we think that it is no longer enough to consider Cloud computing and IoT as two different entities, but we need to change the perspective, especially when looking at how to protect IoT systems. Similarly to other works in the literature, such as \cite{Botta_2014,Diaz_2016,Cook_2018}, we assume a picture of IoT in which Cloud computing and end devices are the two tight layers constituting a broader Internet of Things.
In this new setting, IoT cannot disregard Cloud computing, as the Cloud is a core component of the overall IoT architecture, rather than an external entity.  Note that the viceversa is not true, as the Cloud was not originally thought for IoT devices and it has been widely studied as a stand-alone paradigm.

\begin{figure*}[tb]
	\centering
	\includegraphics[clip, width=0.85\textwidth]{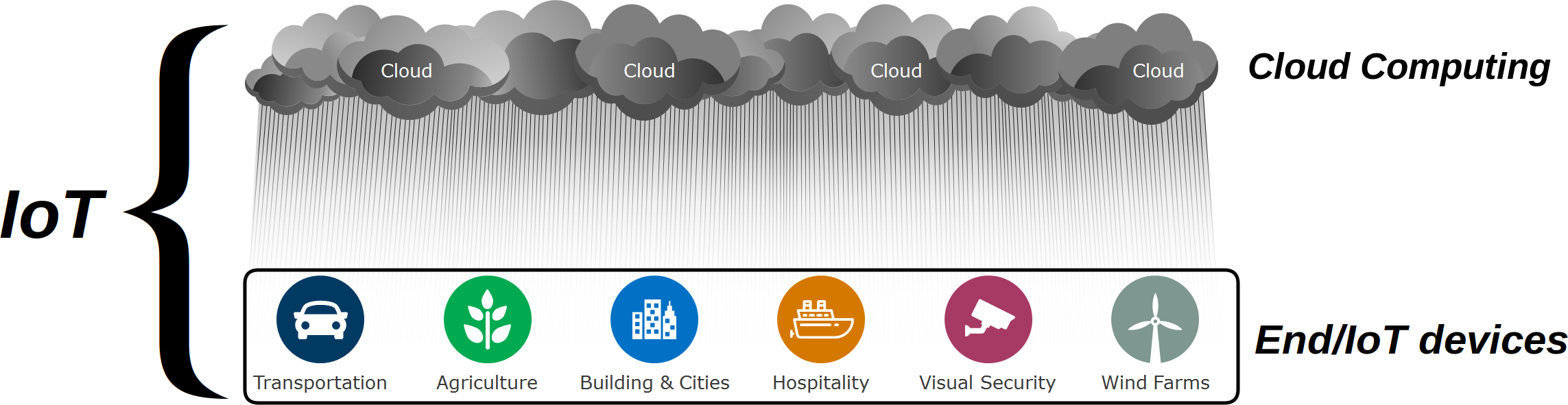}
	\caption{A broader definition of IoT (adapted from \cite{OpenFog_2017_Architecture}): a two layered architecture in which End/IoT devices strongly rely on the Cloud}
	\label{fig:cloud_IoT}
\end{figure*}

From a security perspective, this vision of Cloud computing as a key component of the IoT architecture implies that all security issues that the Cloud drags on need to be analyzed and addressed when referring to IoT security. The result, depicted in Fig.\ref{fig:cloud_IoT}, is a metaphoric rainstorm of cyber-security issues potentially affecting every context of the current and future society. For this reason, we strongly believe that a clear and detailed analysis of the security issues of the \quotes{clouds} is essential to improve the security on the \quotes{ground}.

%



\subsection{Contribution and Outline of the Paper}
This paper aims at providing an up-to-date and well-structured survey of the security issues of Cloud computing in the era of the IoT revolution. Hence, we do not aim at proposing yet another survey of security issues of Cloud computing as a stand-alone paradigm, but we aim at discussing security issues of the Cloud when considered as a core component of the broader IoT architecture. For this purpose, we use a structured approach. First, we distinguish security issues specific of Cloud computing from issues not strictly related to the Cloud but still having an impact on the overall IoT architecture (depicted in Fig.~\ref{fig:cloud_IoT}). Then, we classify both types of issues according to two different angles: the affected Cloud architectural layer and the impacted security property (in terms of confidentiality, integrity, availability). We believe that this classification is vital to understand security issues of Cloud computing, having a clear picture of where issues occur and what their potential impact is. Since there is no IoT without Cloud, we cannot secure IoT without securing the Cloud.

In summary, the contribution of the paper is twofold:
\begin{itemize}
	\item We provide a novel Cloud-centered perspective of IoT security. As already mentioned, Cloud computing has become of paramount importance for Internet of Things. Nevertheless, most of the works related to IoT security focus on the security of end devices. In this paper, we fill this gap providing an analysis of Cloud security issues and how they affect IoT security.
	\item We propose and discuss a structured classification of Cloud computing security issues:
	differently from other works, security issues associated to Cloud computing will be classified according to different layers. First, we distinguish between Cloud-specific security issues and other issues non strictly related to the Cloud  but still important in the IoT context. Then, for each layer of the Cloud architecture, we investigate security properties affected by each issue. This contribution aims at giving a clear overall picture of all aspects of Cloud security.
\end{itemize}


\textit{Outline of the Paper}.
The rest of this work is organized as follows. Section~\ref{sec:related_work} motivates our research by reviewing similar efforts and by comparing them with the rationale behind our manuscript. Section~\ref{sec:cloud_computing} gives basic notions on Cloud computing. Section~\ref{sec:taxonomy} describes the methodology adopted in our research, which is of key importance in order to understand the classification proposed in the paper. In particular, it first depicts the assumed reference architecture. Then, it explains how the classification has been structured. Sections~\ref{sec:specific_cloud_issues} and~\ref{sec:generic_cloud_issues} discuss the Cloud-specific security issues and the Cloud-generic security issues, respectively. Finally, Section~\ref{sec:conclusion} wraps up and concludes the work.

\section{Motivation}\label{sec:related_work}
In this section we review relevant works that have inspired and motivated our research. To this aim, we focus the discussion on how our contribution extends and complements the literature.

Subashini and Kavitha \cite{SUBASHINI_2011} group security issues in relation to the service model they affect, having a focus on the Software as a Service (SaaS) one. For each service model, the authors report different categories of security issues without a clear classification criteria.
The result is a mixture of categories often overlapped with each other. We claim that this lack of separation between classes, along with the intrinsic complexity of the Cloud, does not allow the reader to develop a clear picture of where issues occur within the Cloud architecture and what security property they affect. 

Grobauer et al. \cite{Grobauer_2011} are the first authors proposing a differentiation between specific and general security issues of the Cloud. They focus on Cloud-specific issues and classify them in relation to the architectural level they occur. However, no focus is placed on the security property each issue affects.

Similarly, Modi et al. \cite{Modi_2013} classify security issues based on a Cloud architecture that is alike to the one used in this paper. However, they do not specify which security property is affected by each issue.

Singh et al. \cite{Singh_2016} group security issues in relation to different categories whose choice is unclear. This makes difficult for the reader to understand how the different categories are related and consequently it complicates the comprehension of security issues. However, some of the identified threats are contextualized with the security attribute they compromise.

Fernandes et al. \cite{Fernandes_2014} produce one of the most comprehensive surveys on Cloud computing security issues. They identify a large number of security issues and group them based on a taxonomy that is clearly defined. Nevertheless, they do not specify which security property is affected by each issue.

Singh and Chatterjee \cite{Singh_2017_cloud} extend the work of Fernandes et al. to include possible solutions to the identified problems.

Xiao and Xiao \cite{Xiao_2013} propose to classify security issues in relation to the properties they affect. However, they identify only a small subset of threats together with a list of possible solutions.


Instead of classifying Cloud security issues at a fine-grained level, Ardagna et al. \cite{Ardagna_2015} choose to classify literature works in relation to the security property affected by the issues considered in such works. However, this coarse-grained approach does not allow to achieve the desired level of detail. Indeed, since many of the classified works do not specify the impact of each issue, the approach used by Ardagna et al. \cite{Ardagna_2015} does not help the understanding of what security property is affected by each security issue.

Hashizume et al \cite{Hashizume_2013} present a categorization of security issues focusing on a service model perspective while distinguishing between threats and vulnerabilities.

To the best of our knowledge, there is no work in the literature proposing a structured classification of Cloud computing security issues in the IoT context. This  represents the main motivation behind this paper.

\section{Background: Cloud Computing Paradigm}  \label{sec:cloud_computing}
Nowadays, Cloud computing is a well-known paradigm. However, for the sake of readability and self-containment of the paper, we consider relevant to recap basic notions 
of Cloud computing. This also allows to define a common terminology that is going to be used throughout the rest of this paper. For these reasons, background notions about Cloud computing are provided in this section.


NIST \cite{NIST2011_Def} defines Cloud computing as \quotes{a model for enabling ubiquitous, convenient, on-demand network access to a shared pool of configurable computing resources (e.g., networks, servers, storage, applications, and services) that can be rapidly provisioned and released with minimal management effort or service provider interaction}.
\begin{figure*}[tb]
	\centering
	\includegraphics[trim=0cm 0cm 0cm 0cm, clip, width=0.75\textwidth]{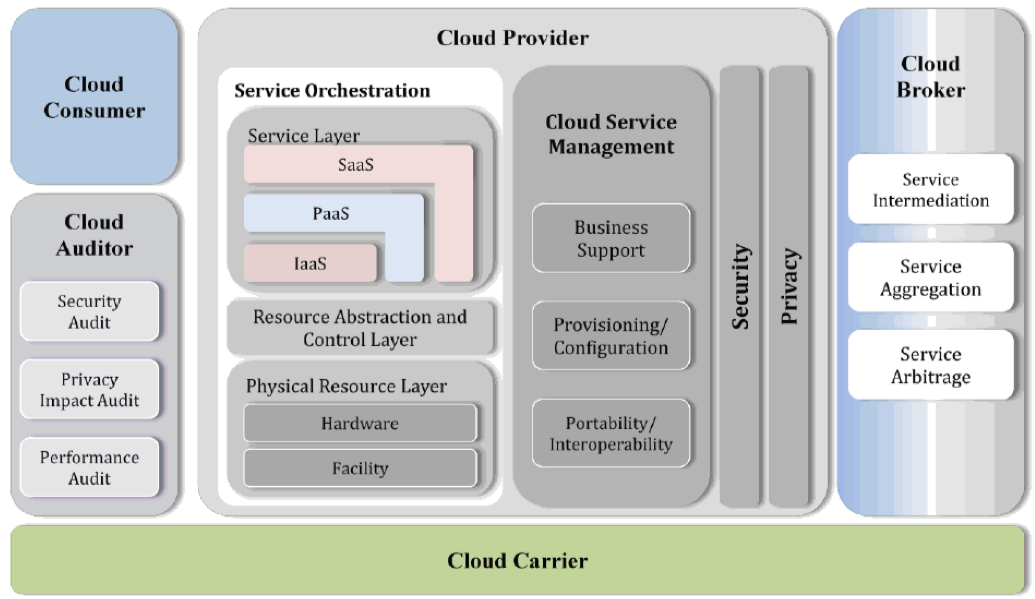}
	\caption{NIST Cloud computing reference architecture (source \cite{NIST2011_Arch})}
	\label{fig:cloud_architecture}
\end{figure*}
%

Figure~\ref{fig:cloud_architecture} depicts the NIST Cloud computing reference architecture \cite{NIST2011_Arch}. It provides a high-level overview of the Cloud and identifies the main actors and their role in Cloud computing. Each actor is an entity, i.e. a person or an organization, that either takes part in a transaction/process or performs some tasks in Cloud computing. There are five main actors: 
\begin{itemize}
	\item \textit{Cloud Provider}: an entity that provides a service to interested parties;
	\item \textit{Cloud Consumer}: an entity that uses a service from, and has a business relationship with, one or more \textit{Cloud providers};
	\item \textit{Cloud Broker}: an entity that mediates affairs between \textit{Cloud providers} and \textit{Cloud consumers}, and that manages the use, performance, and delivery of Cloud services;
	\item \textit{Cloud Carrier}: an intermediary that supplies connectivity and delivery of Cloud services from \textit{Cloud providers} to \textit{Cloud consumers};
	\item \textit{Cloud Auditor}: a party that conducts independent assessments of the Cloud infrastructure, including services, information systems operations, performances, and security of the Cloud implementation.
\end{itemize}

In terms of interactions, there are several possible scenarios \cite{NIST2011_Arch}. Generally, a Cloud consumer may request a Cloud service from a Cloud provider, either directly or via a Cloud broker. A Cloud auditor conducts independent audits and may contact other actors to collect the necessary information.

The NIST defines the Cloud by means of five essential characteristics, three service models, and four deployment models \cite{NIST2011_Def}.

\subsection{Essential Characteristics} \label{subsec:characteristics}
The essential characteristics of Cloud computing can be summarized as follows \cite{NIST2011_Def}:
\begin{itemize}
	\item \textit{On-demand self-service}: computing capabilities can be provided automatically when needed, without requiring any human interaction between consumer and service provider;
	\item \textit{Broad network access}: computing capabilities are available  over the network and accessible through several mechanisms which are disposable for a wide range of client platforms (e.g., workstations, laptops, and mobile devices);
	\item	\textit{Resource pooling}: computing resources are pooled to accommodate multiple consumers, dynamically allocating and deallocating them according to consumer demand. In addition, the provider resources are location independent, i.e. the consumer does not have any knowledge or control of their exact location;
	\item \textit{Rapid elasticity}: computing capabilities can flexibly be provided and released to scale in and out according to the demand. As a result, the consumer has the perception of unlimited, and always adequate, computing capabilities;
	\item \textit{Measured service}: resource usage can be monitored and reported according to the type of service offered. This is particularly relevant in charge-per-use, or pay-per-user, services because it grants a great transparency between the provider and the consumer of the service.
\end{itemize}

A \textit{Cloud infrastructure} is a collection of hardware and software that empowers the aforementioned essential characteristics of Cloud computing.

\subsection{Service Models} \label{subsec:service_models}
The three main types of service models used in Cloud computing are described below \cite{NIST2011_Def}:
\begin{itemize}
	\item \textit{Infrastructure as a Service (\emph{IaaS})}: processing, storage, networks, and other fundamental computing resources (both software and hardware) are provided to the consumer. The consumer can run and deploy any software and can control operating systems, storage, and deployed applications. The consumer does not control or manage the underlying Cloud infrastructure;
	\item \textit{Platform as a Service (\emph{PaaS})}: the consumer is provided with a whole development stack that can be used to develop and deploy new applications. The development stack includes programming languages, libraries, services, and tools that are supported by the provider. The consumer controls both deployed applications and possible configuration settings for the applications environment. The consumer does not control or manage the underlying Cloud infrastructure, operating systems, and storage;
	\item \textit{Software as a Service (\emph{SaaS})}: the consumer can use the applications offered by the provider, running on the Cloud infrastructure.  The consumer does not control or manage the underlying Cloud infrastructure, operating systems, storage, and individual applications capabilities.
\end{itemize}

\begin{figure}[b]
	\centering
	\includegraphics[trim=0cm 0cm 0cm 0cm, clip, width=0.480\textwidth]{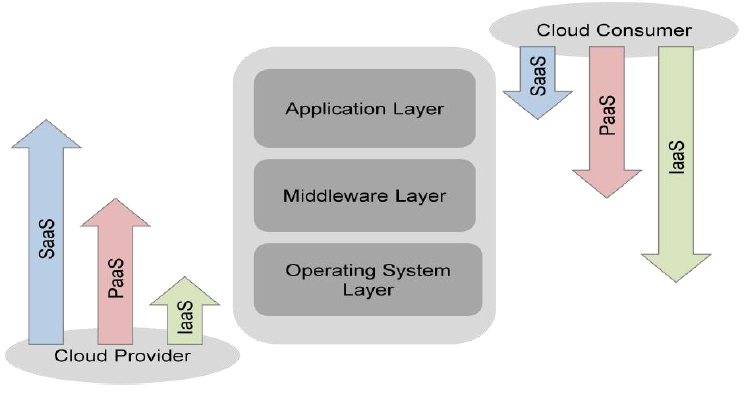}
	\caption{Scope of control between provider and consumer (source \cite{NIST2011_Arch})}
	\label{fig:scope_controls}
\end{figure}

In all the service models, Cloud provider and Cloud consumer share the control of the Cloud system. However, as shown in Fig.~\ref{fig:scope_controls}, each service model implies a  different degree of control over the computational resources for each party, thus different responsibilities \cite{NIST2011_Arch}.

\subsection{Deployment Models} \label{subsec:deployment_models}
The four main models used for the deployment of Cloud computing are discussed below \cite{NIST2011_Def}:
\begin{itemize}
	\item \textit{Private Cloud}: the Cloud infrastructure is provided for the exclusive use of a single organization. The organization can include different consumers (e.g., business units);
	\item \textit{Community Cloud}: the Cloud infrastructure is provisioned for the exclusive use of organizations with shared concerns, such as security requirements, policy, and mission. Each organization can include multiple consumers;
	\item \textit{Public Cloud}: the Cloud infrastructure is provided for the open use by the general public over the Internet. It is ideal either for small to medium size businesses, or for single customers;
	\item	\textit{Hybrid Cloud}: the Cloud infrastructure is a merge of two or more infrastructures deployed with different models (private, community, or public). Each Cloud infrastructure remains a unique entity, but it is bound together with the others by standardized or proprietary technologies enabling portability.
\end{itemize}

In all the aforementioned models, the Cloud infrastructure may be owned, managed, and operated by one or more consumer organizations (if any), a third party organization (e.g., business organization, academic organization, or government organization), or any combination of them.

\section{Methodology} \label{sec:taxonomy}

In this section, we introduce the methodology adopted to classify security issues. First, we describe the simplified Cloud architecture that we use as a reference. Then, we explain how the classification is organized.

\subsection{Reference Architecture}\label{subsection:SimplifiedArchitecture}
\begin{figure}[tb]
	\centering
	\includegraphics[scale=0.3]{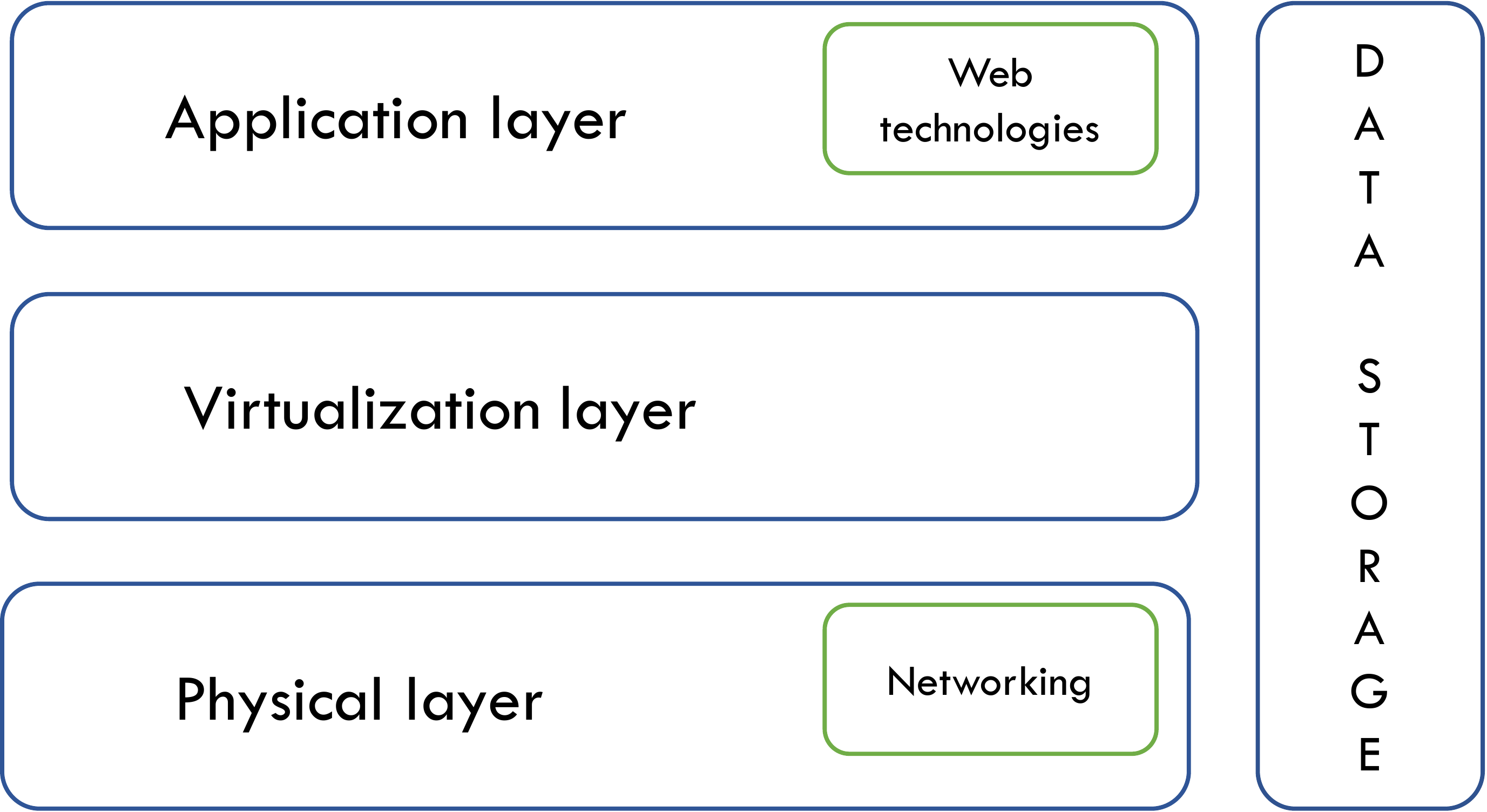}
	\caption{Simplified Cloud reference architecture}
	\label{fig:cloud_reference_architecture}
\end{figure}

Cloud computing is one of the most complex computing
paradigm existing today. For this reason, it is essential to take apart irrelevant details when it comes to classify its security issues. To reach this objective, we introduce a simplified architecture of the Cloud infrastructure, which is depicted in Fig.~\ref{fig:cloud_reference_architecture}. This architecture is an abstraction of the architecture proposed in \cite{Modi_2013} and it is simplified to such an extent that Cloud computing is considered as composed of four main layers: \textit{physical layer}, \textit{virtualization layer}, \textit{application layer}, and \textit{data storage}.
%

The key components we consider at the physical layer are computational, storage, and networking resources. However, since security issues of physical resources are beyond the purposes of this work, at this layer we only consider \textit{network security issues}.

In the virtualization layer, we locate Virtual Machines (VM), Virtual Machine Monitors (VMM), virtual networks, and all the infrastructure directly or indirectly supporting virtualization (e.g., mechanisms enabling virtual machine migration, management of VMs, and so on).

We consider all the remaining software as part of the application layer: specific applications, APIs, tools, middlewares, management services, monitoring systems, load balancing systems, and others. Further, all software (above the virtualization level) used to build PaaS and SaaS Cloud implementations is considered part of the application level. Hence, in this respect, we consider PaaS and SaaS as parts of the application level. Indeed, we see them just as any other application offering some special type of services.

Finally, we consider data storage services as part of all the layers of the architecture, therefore, they are treated alongside the other layers.

\subsection{Structured Classification}
\begin{figure*}[tb]
	\centering
	\includegraphics[scale=0.3]{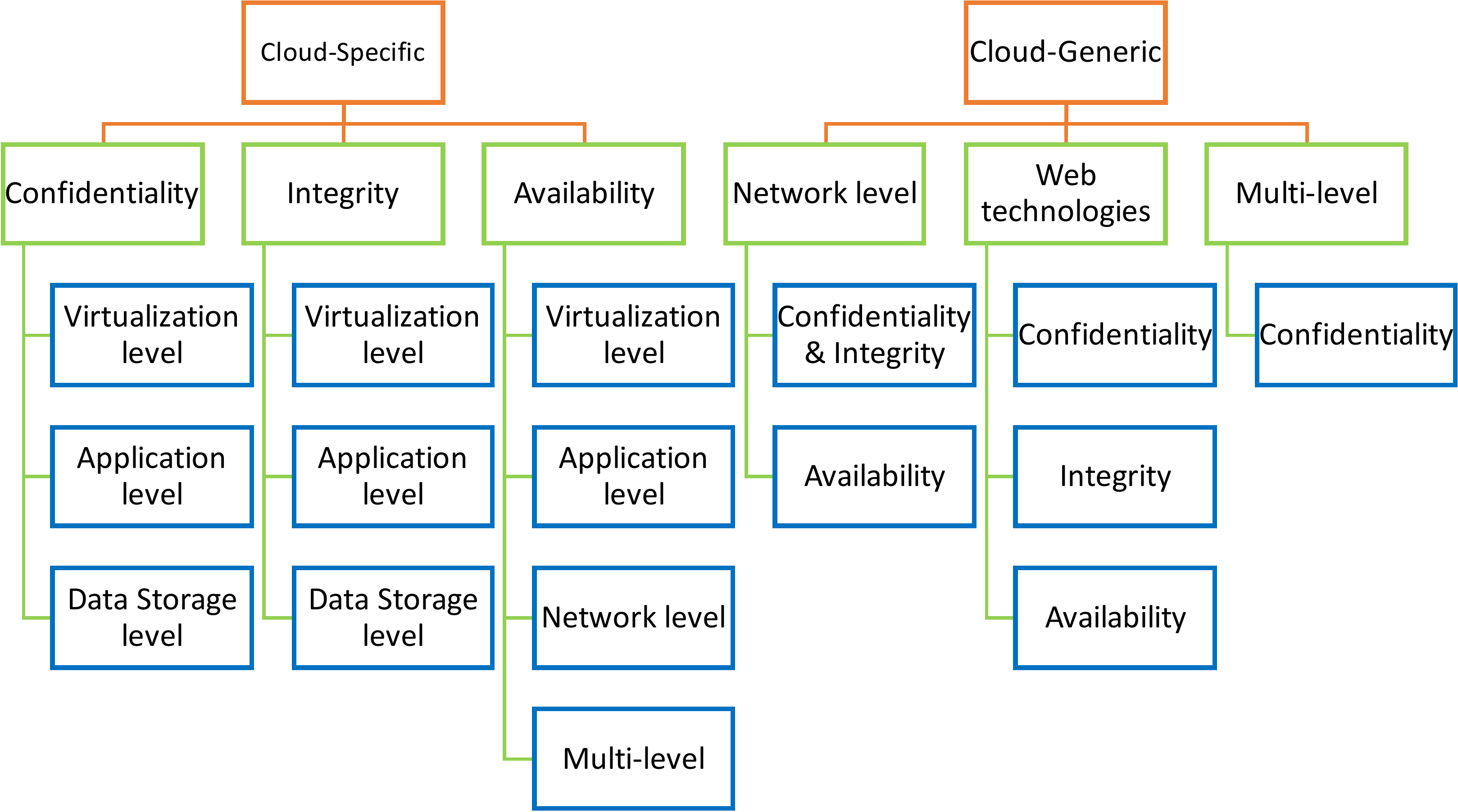}
	\caption{Classification of Cloud security issues}
	\label{fig:taxonomy}
\end{figure*}

In this section, we describe how our reference architecture is adopted to classify Cloud security issues. The overall classification is depicted in Fig.~\ref{fig:taxonomy}.

Firstly, we separate Cloud-specific security issues from Cloud-generic ones. Details about the criteria used for performing such distinction are provided in Sec.~\ref{sec:specific_cloud_issues}. In short, many security issues of the Cloud exist also in other paradigms, since rooted in common technologies employed to build distributed systems. Thus, we distinguish between issues that we consider specific of the Cloud environment and other common security issues not strictly related to the Cloud but still having an impact on the overall IoT architecture (depicted in Fig.~\ref{fig:cloud_IoT}). However, even if we also present a subset of Cloud-generic issues, our main focus is on Cloud-specific ones.

Secondly, security issues are further classified from two different perspectives: the Cloud architectural level at which they occur and the security property they affect. In other words, given a certain level $x$ of the Cloud reference architecture and a certain security property $y$, the following questions are answered: \textit{1) What are the security problems at level $x$ of the Cloud architecture?}, \textit{2) How do they affect property $y$?} In answering these questions, the security properties we consider are the well-known confidentiality, integrity, and availability (CIA). We have decided to stick only with these security properties to keep the scope of the paper well focused and manageable in terms of literature and analysis. However, the same methodology can be applied to and iterated with other security properties (e.g., authenticity and accountability).

\begin{figure}[tb]
	\centering
	\includegraphics[scale=0.6]{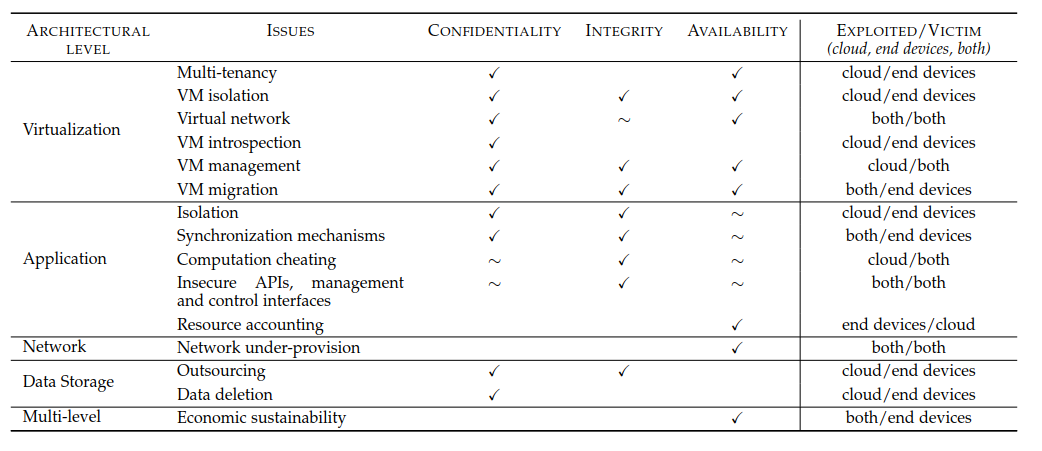}
		\caption[caption]{Summary of Cloud-specific issues\\
				\quotes{\checkmark}: existence of literature works indicating that the issue affects the property.
				\quotes{$\sim$}: despite we found no evidence in the literature, we believe that the issue might affect the property.
				\textsc{Exploited/Victim}: how parties of the IoT architecture (Figure \ref{fig:cloud_IoT}) are affected from the issue.}
	\label{tab:cloud-specific issues}
\end{figure}
\begin{figure}[tb]
	\centering
	\includegraphics[scale=0.6 ]{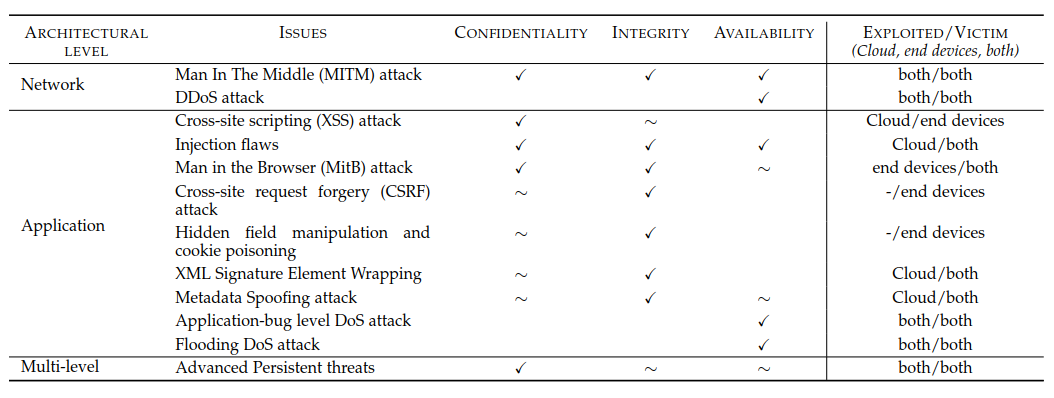}
		\caption[caption]{Summary of Cloud-generic issues\\
			\quotes{\checkmark}: existence of literature works indicating that the issue affects the property.
			\quotes{$\sim$}: despite we found no evidence in the literature, we believe that the issue might affect the property.
			\textsc{Exploited/Victim}: how parties of the IoT architecture (Figure \ref{fig:cloud_IoT}) are affected from the issue.}
		\label{tab:cloud-generic issues}
\end{figure}

The classification resulting from the analysis described in Sec.~\ref{sec:specific_cloud_issues} and Sec.~\ref{sec:generic_cloud_issues} is depicted at the end of the paper in Table~\ref{tab:cloud-specific issues} and Table~\ref{tab:cloud-generic issues}, respectively. These tables show each issue in relation to the architectural level it occurs and the CIA property it affects. For each cell of the table (associated to a specific pair: issue, security property) a mark is applied according to the following rules:
\begin{itemize}
	\item \quotes{\checkmark}: it is placed if we found a literature work describing an attack affecting the corresponding security property, or if we found a literature work stating that the issue might affect the corresponding property;
	\item \quotes{$\sim$}: it is placed if, although the previous condition is not verified, we believe that the given issue might allow to compromise the corresponding security property;
	\item an empty cell, if the previous conditions do not hold.
\end{itemize}
Moreover, in the last column of the tables, we highlight the relation of each security issue with Cloud and end devices. In details, we indicate which party can be exploited because of the specific security issue, and which party might be the victim of an attack perpetrated exploiting that issue. If neither the Cloud nor end devices are involved, we draw a \quotes{-}.

%

\section{Cloud-specific security issues} \label{sec:specific_cloud_issues}
In this section, we present security issues peculiar for Cloud computing.
Inspired by the work in \cite{Grobauer_2011}, we consider as Cloud-specific issues all those problems that are rooted in at least one of the essential Cloud characteristics defined by NIST. Please consider that, according to such definitions, 
network-level and web-technologies issues (discussed in Sec.~\ref{sec:generic_cloud_issues}) should be considered specific for the Cloud. However, since these security issues are also really common in a number of distributed paradigms, we have decided to consider them as Cloud-generic security issues and to not discuss them in this section. 

In the following, we present Cloud-specific security issues based on a two layer classification. First, we classify security issues based on what CIA propriety they affect. Then, for each property, the issues are further organized in relation to the Cloud architectural level they affect.

\subsection{Confidentiality}\label{subsubsection:conf}
According to \cite{GlossaryOpenFog_2018}, confidentiality is the \quotes{property that information is not made available or disclosed to unauthorized individuals, entity or processes}. Hence, it is the property indicating absence of unauthorized disclosure of information and data \cite{GoodricTamassia_2011}.
We present a classification of security issues that can impair data confidentiality. Each class of our classification is a component of the Cloud architecture (defined in Section \ref{subsection:SimplifiedArchitecture}) while the entries of each class are the security issues that have causes rooted in that specific level.
	\subsubsection{Virtualization level issues}
		Virtualization technology is one of the key enabler of Cloud computing. However, this additional abstraction layer has severe security repercussions. In the following paragraphs we report key security issues caused by this layer and capable of compromising data confidentiality.
		\paragraph{Multi-tenancy issues} Virtualization technology allows to develop a multi-tenant environment in which virtual machines operate sharing communal hardware resources. The placement of different users on the same platform is what enables new types of attacks on data confidentiality. In \cite{Ristenpart_2009}, the authors describe how they were able to exploit several characteristics of Amazon Elastic Compute Cloud (EC2) in such a way to have their own virtual machine co-resident (i.e. on the same physical-machine) with that of a victim. Once co-residence is reached, an attacker has the unprecedented possibility of performing several types of side-channel attacks in such a way to extract confidential information from users who are sharing the same machine with the attacker. In \cite{Ristenpart_2009}, it is shown that, by means of cache measurements, an attacker can perform: keystroke timing attack, traffic rates estimation of victim's web servers and even co-residence detection. Moreover, side channel attacks affecting cryptographic implementations have been reported in \cite{Zhang_2012,Yarom_2014,Liu_2015,Irazoqui_2014}. The work in \cite{Suzaki_2011} shows the possibility to exploit memory deduplication issues for performing another type of cross-VM side channel attack.  Further, the recent vulnerabilities Meltdown\cite{Meltdown} and Spectre \cite{Spectre} have demonstrated that not only memory-based side channel attacks are possible, but that even processor vulnerabilities can be exploited to perform attacks capable of breaking any security assumption and allowing other co-resident VMs to access confidential information belonging to other users.

		\paragraph{VM Isolation issues} According to \cite{Singh_2016}, virtual machine isolation is the principal factor that can lead to cross-VM data leakage. Virtualization is based on the hypervisor ability of isolating VMs from each other. However, due to several reasons (e.g. misconfiguration, design and implementation bugs) an attacker can compromise the hypervisor, evade from isolation and potentially take over all the other guests \cite{WangWu_2012}. We refer to such situation as virtual machine escape \cite{Riddle_2015}. Escaped VMs can access data and information belonging to other VMs\cite{Studina_2012}, resulting in paramount confidentiality issues. Appropriate security mechanisms are therefore required for isolating virtual machines from each other and hence preventing data leakage. Some possible techniques for isolation enforcement are described in \cite{Studina_2012}, while in \cite{Riddle_2015}, techniques for providing integrity of VMM are reported.

		\paragraph{Virtual network issues} According to \cite{Li_2012} and \cite{AlMorsy_2011}, not only virtual machine isolation is needed but also isolation of virtual networks is required. Therefore, virtual networks are another source of vulnerability for confidentiality and, as such, need to be protected. Even though some traditional controls (such as virtual local-area networks and firewalls) have been proven to be less effective in virtual networks \cite{Vaquero_2011}, in \cite{Li_2012}, the authors, propose to implement traditional network security solutions into virtual environments. Typical confidentiality threats that can affect virtual networks are sniffing and spoofing attacks \cite{Wu_2010}. Moreover, if from a user's prospective the virtual network is a private network, it might in reality rely on a public infrastructure and therefore appropriate protections to secure communications are needed \cite{Schoo_2011}.

		A novel type of attack that exploits virtual networks as a cornerstone for subsequently compromising the whole Cloud system is the \quotes{virtual switch attacker model for packet-parsing} (vAMP attack)\cite{Thimmaraju_2017}. This attack exploits vulnerabilities of specific packet parsing systems deployed in virtual switches for generating a series of attacks that eventually allow to take control of the entire Cloud system.

		\paragraph{Virtual Machine introspection issues}  Different literature works, such as\cite{LombardiDiPietro_2010},  \cite{Li_2012} and \cite{Jiang_2007}, propose to use the hypervisor for monitoring virtual machines with the objective of  preventing or discovering attacks on integrity of guest systems. If from the one hand this kind of approach presents important advantages, on the other hand it also highlights the possibility for the Cloud provider or malicious insiders (or even for an external attacker able to take control of the hosting platform) to break users' confidentiality by exploiting virtual machine introspection. This problem is linked to the more general and emblematic question of deciding whether the Cloud provider and the infrastructure he provides, should be considered trusted or not; which is a typical problem of every scenario in which outsourcing is present. It is worth noting that, in case the Cloud provider is considered trusted, the Cloud infrastructure might also play a key role in solving many of the existing security issues \cite{Aikat_2017}.

		An example, of attack that can allow a malicious insider to exploit virtual machine introspection is described in \cite{Rocha_2013}.

		For the sake of completeness, it should also be mentioned that attacks targeting virtual machine introspection mechanisms have been reported in literature. An example of such attack is Direct Kernel Structure Manipulation (DKSM)\cite{Bahram_2010}.

		\paragraph{VM management issues} VM image cloning enables Cloud providers to supply on-demand services to their clients. Cloned VMs can be moved on different servers in relation to clients' needs but this also makes clients unaware of how many VMs copies exist, were these are specifically located and who is possessing them. Such availability, allows a malicious insider to exploit one of the existing VM copies to attempt breaking the VM password and gain access to all the information saved into the VM \cite{Duncan_2012} while leaving the owner unaware of such situation.

		VM image sharing is another key service enabled by VM image cloning. VM image sharing is one of the Cloud foundations\cite{Wei_2009}, however both the VM image publisher and the retriever are subject to confidentiality concerns\cite{Balduzzi_2012}. Indeed, by publishing an image, the publisher may release his own confidential information, while, on the other side, user's data confidentiality can be compromised by shared malicious images, for instance, they can contain back-doors for silently access confidential data\cite{Grobauer_2011,Modi_2013_IDS}. Moreover, VM image sharing makes also possible for attackers to rent cloned VMs with the only purpose of analyzing their content and therefore to identify possible vulnerabilities that could be exploited in future attacks.  

		\paragraph{VM migration issues} Virtual Machine migration allows to transfer running VMs from one host to another in a transparent fashion for the final user \cite{Aiash_2014,Yeh_2016}. The Cloud advantages of using such mechanisms are obvious, just to name a few: it enables load-balance when hosts are overloaded, it allows to reduce costs through VMs consolidation  and improves the overall manageability of the system \cite{Rakotondravony_2017,Aiash_2014,Yeh_2016}. However, protocols used in implementing live migration have to be secured, since, if control messages and VM to be migrated are not encrypted, then common attacks on confidentiality (such as eavesdropping for sensitive data and passwords) might be easily performed \cite{Aiash_2014,Jyoti_2012}.

	\subsubsection{Application level issues}
		We are now going to consider which are the Cloud issues for confidentiality whose causes are rooted at the application level.
		According to our reference architecture (defined in Section \ref{subsection:SimplifiedArchitecture}), every software deployed on top of the virtualization layer has been considered part of the application level. Since we consider PaaS and SaaS systems special type of application-level services, these are considered part of this level too. We remind to the reader that even if we consider web-related issues part of the application layer, they are not specifically related only to the Cloud but common of any distributed system and for this reason these are discussed in Sec.~\ref{sec:generic_cloud_issues}.

		\paragraph{Isolation issues} users of PaaS systems can develop and run their own applications on platforms provided by Cloud providers. These platforms allow applications developed by different users to share communal libraries and supporting services \cite{Rodero_2012}. Even if the platform (or container system) can be a proper Operating System, in most cases it is a Virtual Platform (e.g. Java or .Net) \cite{Rodero_2012}. Irrespectively from the specific type of implementation, a common concern of PaaS systems is to ensure that isolation of tenants is properly implemented and that an application can not explore or modify other data and applications. The work in \cite{Rodero_2012}, presents a panoramic of isolation issues that could have arisen when Java or .Net technologies were used to create PaaS implementations. 
		It should at this point be noted that PaaS implementations vary deeply from provider to provider \cite{Linthicum_2017} and therefore Java or .Net may neither be actually used in the majority of systems nor present these flaws anymore. Nevertheless, the issues which are reported in this paper have the goal to identify the Cloud attack surface and make aware of the possible vulnerabilities that can affect the Cloud environment, independently from the specific technology used but in relation to what can possibly go wrong and become an issue for the Cloud environment.

		Within SaaS models, multitenancy is present also at the application level. In \cite{AlMorsy_2011}, the authors describe how multitenancy can be implemented for allowing the same application to be shared among different users. As result of multitenancy at application level, data of different users are stored in common structures \cite{SUBASHINI_2011} which enables malicious tenants to exploit applications loop holes, masked code injection or security misconfigurations to sneak into other users data \cite{SUBASHINI_2011},\cite{AlMorsy_2011}.

		\paragraph{Synchronization mechanisms issues} synchronization mechanisms are common in Cloud storage SaaS implementations \cite{Nakouri_2017}. When modification of files are performed on a local device, such mechanisms allow to propagate updates to all other devices interested in those files \cite{Nakouri_2017}. These mechanisms are typically implemented by the use of tokens which have been shown to introduce new vulnerabilities that can allow to perform data exfiltration\cite{Nakouri_2017,Liang_2017}. An example of attack exploiting such vulnerability is the Man in the Cloud (MitC) attack \cite{Liang_2017}.

	\subsubsection{Data Storage level issues}
		In the following paragraphs we are going to report some confidentiality issues that, despite being specific of the Cloud, are not strictly related to a specific level of the Cloud architecture but that embrace more than one level of the architecture.
		\paragraph{Outsourcing issues} Applications deployed on the Cloud have to be remotely accessed by users who, depending on the type of application and elaboration needed, may be requested to outsource private and confidential information. The immediate consequence of outsourcing is loss of control which impede the owner of outsourced data to directly dispose and control them as he prefers, making it difficult to protect confidentiality with traditional methods \cite{Xiao_2013}. To understand the reasons behind such difficulty it is paramount to distinguish between applications offering storage services and applications offering some type of remote elaborations. In both cases, it is legitimate to assume that the service provider will implement access policies and security mechanisms for protecting users' data \cite{Samarati_2016} but it also implies that he is in the perfect position to access such data and therefore break users' data confidentiality. However, while in the former case users can easily prevent such situation by encrypting data before their are stored in the Cloud (which could also make it much more secure than storing them unencrypted in private data centers \cite{Zhou_2010}), in the latter case, the possibility to protect confidentiality by means of traditional encryption schema is not feasible due to the service provider need of performing elaborations\cite{Chen_2012}. Nevertheless, plain text data should be avoided in order to prevent Cloud providers from accessing information which, due to the lack of control, could even be stored or transmitted to third parties and be used for other purposes (there are great examples in literature demonstrating how such situations can produce unwanted consequences; some of these threats, which are also related to multi location, can be found in \cite{Zhou_2010}). If we consider that Cloud applications take advantage of composite request processing\cite{Helland_2013}, which allows service providers itself to outsource part of the computation, it is clear that the confidentiality risks are even more higher. Full homomorphic encryption could be the solution to alleviate confidentiality concerns of outsourced data but according to \cite{Martins_2017} and \cite{Rong_2013} this approach is neither efficient yet nor adequate for general purpose elaborations.

		In some cases, even applications offering a pure storage service may still require some amount of computations on encrypted data (for instance, content research may be required for enabling fine-grained retrieval) \cite{Ren_2012}. To face this necessity, confidentiality-preserving query evaluation approaches are reported in \cite{DeCapitani_2018}, but, similarly to the case of homomorphic encryption, they only support partial query execution. Moreover, even if encryption or fragmentation techniques are used to protect confidentiality of data, it may also be required to hide information about which data is accessed (access confidentiality) together with the patterns exhibited in accessing such data (pattern confidentiality) \cite{Samarati_2016,Tari_2014}. Indeed, in \cite{Islam_2012} it is demonstrated that lacks in protecting such information can result in contents disclosures.

		In case that data are remotely elaborated on the Cloud by means of programs written by the owner of such data (which is typically the case for IaaS and PaaS services), to protect confidentiality and integrity from an untrusted Cloud provider, solutions relying on Intel software guard extensions (SGX) have recently been proposed \cite{Baumann_2015}. SGX features, allow processors to instantiate secure memory regions which are protected from hardware attacks or malicious privileged code \cite{Baumann_2015}. This capability could therefore be used for executing programs in the Cloud with a similar level of security to the one in which programs are executed on hardware resources belonging and controlled by the owner of data\cite{Baumann_2015}.

		\paragraph{Data deletion issues} data deletion need special attention since if it is not correctly performed it leads to greater confidentiality threats. From the one hand, even if the delete operation has been correctly performed, integrity of the operation can indirectly be breached due data recovery vulnerabilities \cite{Grobauer_2011}. An example of such situation arises due to the physical features of storage devices which can allow to restore original data \cite{Chen_2012} even if the delete operation has actually been performed at software level. On top of these cases, the service provider may directly impact on the integrity of the delete operation by incorrectly performing such operation (for instance due to not properly taking into account data replication) \cite{Pearson_2013} or even by not performing it at all.

\subsection{Integrity}\label{subsubsection:int}
Integrity is the \quotes{assurance that the information is authentic, complete and can be relied upon to be sufficiently accurate for its purpose. It refers to whether the information is correct and can be trusted and relied upon} \cite{GlossaryOpenFog_2018}. We extend such definition to embrace also integrity of computations. This implies that integrity is also about guaranteeing that information resulting from computations is authentic, complete and can be relied upon.

The same classification of security issues that has been previously performed in relation to confidentiality is going to be repeated for integrity issues.

	\subsubsection{Virtualization level issues}
	   In the following paragraphs security issues rooted in the virtualization layer and with the potential to impact integrity of data are presented.
		\paragraph{VM isolation issues} at this level, virtual machine escaping is the way in which data and software integrity can be attacked. Indeed, a compromised VMM can threaten integrity of data \cite{Samarati_2016}. More specifically, if a virtual machine is able to escape from isolation and compromise the VMM, it can access memory locations belonging to other users while having the required privileges to write or delete their content \cite{Studina_2012}\cite{WangWu_2012}, in such a way to perform a VM hopping attack \cite{Jasti_2010,Tsai_2012}. The VMM can possibly be attacked through several attack vector: device drivers, VM exit events or hypercalls \cite{Milenkoski_2013}; a throughout list of vulnerabilities typical of common VMMs used to deploy Cloud systems, can be found in \cite{Perez-Botero_2013}.  For this reason, in order to protect users' data integrity it is essential to protect the isolation capabilities and integrity of virtual machine monitors. A list of possible mechanisms to guarantee VMM integrity and enhance isolation is reported in \cite{Studina_2012} and \cite{Riddle_2015}.

		\paragraph{VM management issues} bad management of VM images has negative repercussion on the integrity of the Cloud environment. Indeed, vulnerabilities in the Cloud environment can be introduced by injecting malware into VM images repositories \cite{Rakotondravony_2017}. Thereafter, with lacks of proper VM image management and controls, sporadically running images are in the perfect position to carry worms and compromise integrity of other images while avoiding detection thanks to low activity level \cite{Wei_2009}. Therefore, integrity checks and scans of VM images are required as consequence of VM cloning and sharing. Moreover, such controls are also paramount in relation to the necessity to protect Cloud repositories against the increasing trend of \quotes{bad repositories}, i.e. the use of Cloud repositories as containers of services for illicit activities \cite{Liao_2016}.

		\paragraph{VM migration issues} live virtual machine migration is paramount for Cloud environments, however it needs to be properly implemented from a security perspective (see also Section \ref{subsubsection:conf}). As for integrity, the attack surface of the migration protocol is potentially quite vast \cite{Aiash_2014}: common vulnerabilities may be used to inject malicious code in the programs implementing migration process; if no encryption is used to secure the exchange of messages controlling the transfer, then, messages might be manipulated to impair integrity of the process; moreover, even compromised hosts might be exploited for affecting integrity of the migrated VM once it is moved to a controlled malicious host.


	\subsubsection{Application level issues}
		We are now going to present integrity issues that are rooted at the application layer. We take into account issues affecting integrity of data and elaborations.
		\paragraph{Computation cheating issues} the combination of outsourcing together with the transparency lack in the way Cloud services are implemented, allows service providers to alter the results of computations or even to not perform elaborations in the proper way \cite{Ren_2012}. If at first such situation might seem strange, there are actually several reasons behind it, e.g. driven by the desire to reduce costs, service providers may be tempted to simplify computations when lots of resources are needed \cite{Wang_2011}. Remote computation can be cheated in several ways: elaborations can be performed on partial or not up to date data, they can be performed incorrectly or may even return partial results \cite{Samarati_2016}, \cite{Wei_2014}. Remote computation audit and verifiable computation have therefore been proposed to face this issue. A review of possible solutions trying to address such problem is presented in \cite{Xiao_2013}.

		Computation might also be cheated not because of the service provider but due to specific attacks. An example of such inconvenience is the Cloud malware injection attack. Cloud providers are responsible for redirecting user's requests toward appropriate services capable of satisfying them \cite{Jensen_2009}. An adversary can exploit such situation to create malicious service implementations, add them to the Cloud and trick the Cloud provider to believe that they are real implementation of some services by falsifying metadata descriptors used to identify functionalities offered by applications \cite{Jensen_2009}. This type of attack results in applications integrity breach since from a users prospective the service has not performed as expected.

		\paragraph{Insecure APIs, management and control interfaces} by means of APIs and management interfaces Cloud users can request, monitor and obtain resources dynamically based on their needs, making the Cloud an on-demand self-service platform\cite{Karnwal_2012}. However, since these interfaces are accessible through the internet and because of web vulnerabilities \cite{Pearson_2013}, the risk of unauthorized access is much higher if compared to traditional systems \cite{Grobauer_2011}. It follows that if an attacker is able to gain unauthorized access to the data contained in such interfaces, then he can compromise services and break applications integrity \cite{Ahuja_2012}. Examples of attacks that in the past compromised control interfaces are reported in \cite{Somorovsky_2011}.

		\paragraph{Isolation issues} isolation issues within platforms used to create PaaS systems (see also subseciton \ref{subsubsection:conf}) can affect integrity of data and applications belonging to other tenants\cite{Rodero_2012}.

		\paragraph{Synchronization mechanisms issues} according to \cite{Liang_2017} vulnerabilities in synchronization mechanisms might also be exploited to compromise integrity of data. An example of attack that can allow to achieve this is the Man in the Cloud (MitC) attack (see also subsection \ref{subsubsection:conf}). Integrity of data can be compromised by such attacks since authentication vulnerabilities are exploited. Therefore, once the attacker takes advantage of tokens and authenticates as a different user, then he is able to impair both confidentiality and integrity of all data belonging to that user.

	\subsubsection{Data Storage level issues}
		In the following paragraph we discuss about integrity issues related to the protection of data storage. We have decided to not directly associate these issues to any of the previous levels as we consider data storage related to all levels of our reference architecture and not predominant of any of them.
		\paragraph{Outsourcing issues} as is the case for confidentiality, outsourcing of data is the Cloud feature that arises new integrity challenges. Data integrity can be compromised in several possible ways and reasons: a Cloud service provider, for economical reasons, may delete users' rarely accessed data in order to release storage space that can be sold to other users; even assuming a perfectly behaving provider, malfunctions are still there to compromise data (which is indeed what happened to Amazon S3 some years ago \cite{Cachin_2009}); more in general, external attackers, driven by economical reasons, might compromise data integrity and this might even not be timely discovered by users \cite{Wang_2012} due the Cloud providers' tendency of hiding unpleasant events that could affect their businesses.
		The need for integrity mechanisms is therefore clear. However, due to outsourcing, traditional integrity mechanisms are non applicable in this scenario since they would require the download of outsourced data for allowing local integrity checks to be performed\cite{Wang_2012,Ren_2012}. Indeed, this is unacceptable for efficiency reasons as it would nullify the Cloud advantages (especially in relation to situation where high amounts of data are outsourced). Therefore, remote data integrity checking protocols are required\cite{Syam_2011}. Nevertheless, challenges do exist for the development of such protocols especially in relation to efficiency requirements and the possibility to guarantee integrity of dynamic data (i.e. data that are modified or updated after they have been loaded in the Cloud). For limited resourced clients, the burden of computation and communication imposed by such protocol has to be as limited as possible which has lead to the idea of using protocols based on third parties auditors \cite{Wei_2014}.
		In \cite{Zafar_2017}, an in-depth review of remote data integrity checking protocol is presented with associated issues for their development and possible attacks they may face.

\subsection{Availability}\label{subsubsection:avail}
Availability is the \quotes{assurance that the systems responsible for delivering, storing and processing information are accessible when needed, by those who need them} \cite{GlossaryOpenFog_2018}. Hence, availability is the property indicating the possibility, for authorized users, to access (and modify) data whenever needed \cite{GoodricTamassia_2011}.

This subsection is aimed at presenting availability and performance degradation issues that arise at the different levels of our architecture.

	\subsubsection{Virtualization level issues}
		Virtualization technology introduces new attack vectors that can be exploited to impact on the availability and performances of Cloud systems. In the next paragraphs we seek to report the main issues we have identified in relation to this concern.
		\paragraph{Multi-tenancy issues} according to \cite{Ristenpart_2009}, an attacker can exploit co-residence, and act on shared physical resources, in such a way to perform denial-of-service attacks or cross-VM performance degradation attacks. The possibility to verify co-residence, might also be exploited to provoke changes in resource utilization of co-resident VMs in such a way to make them use less resources (and hence impacting on their availability) and therefore let the attacker gain high resource availability. This attack is known as Resource-Freeing attack \cite{Varadarajan_2012}.
		\paragraph{VM management issues} availability issues may also arise due to bad VM management policy. An example of such eventuality is VM sprawling, which is a situation where the number of hosted virtual machines keep increasing while most of them are idle \cite{Luo_2011}. VM sprawling can also result from specific attacks aiming at discarding confirmation messages generated from the Cloud service to confirm users that their requests of VM execution has been correctly performed. If users do not receive such confirmation messages, they will keep instantiating VMs even if their action has already been performed. This attack leads to the creation of orphan VMs which can degrate performance and eventually  exhaust the pool of resources \cite{Dabrowsk_2011}.
		\paragraph{VM isolation} availability can be compromised by virtual machines breaking out of isolation and being able to either use all host resources or performing a system halt \cite{Studina_2012}.

		Scheduling issues might be exploited to impact on the performance (and also availability) of other VMs.  Indeed, an attacker can manipulate hypervisor scheduling mechanisms in such a way to obtain more resources for his own VM at the expenses of other clients \cite{Zhou_2013}. Such situation, taken to the limit, can lead to starvation of other VMs or, more in general, can degrade services to such an extent of making services deployed within VMs unusable.
		\paragraph{Virtual network issues} according to \cite{Vaquero_2011}, poor scalability of virtual networks  is another factor that can be exploited for a denial of service (DoS) attack.
		\paragraph{VM migration issues} malicious VMs can take advantage of live virtual machine migration to perform DoS attacks or achieve performance degradation. The migrant attack is an example of such type of DoS attack. In a migrant attack, a small set of compromised VMs are coordinated to generate useless resource consumption in order to mislead the Cloud monitoring mechanisms to trigger migration processes \cite{Yeh_2016}. Since live migrations are expensive processes, this allows attacker to waste Cloud resources and degrade performances of other VMs. An equivalent class of DoS attack similar to the previous one, is Cloud-Internal Denial of Service attacks (CIDoS) \cite{Alarifi_2013}.

		Researchers in \cite{Atya_2017} and \cite{Moon_2015}, proposed to use live migration for reducing the time of co-residency among virtual machines and hence prevent side-channel attacks. However, it has been recently shown that it could be possible for an adversary to slow down migration processes and therefore still permit the attackers to perform side-channel information stealing \cite{Atya_2017_USENIX}. In relation to availability, this attack (known as stalling attack) demonstrates the possibility for co-resident adversaries to prevent migrations and hence degrade performances by obstructing the performance gain that would follow from migrations.

		Cloud-Droplet-Freezing (CDF) is another type of DoS attack which is based on the observation that if migrations of VMs are carried on during a flooding attack for the purpose of load-balancing and trying to mitigate the attack, then it might also contribute to increase the overhead for the Cloud and weaken even more its resource availability \cite{Wang_2014}.

	\subsubsection{Application level issues}
		By excluding application layer protocols that support networking (which are not specific of the Cloud, and for this reason discussed in Section \ref{sec:generic_cloud_issues}), at this layer, we have identified only one relevant Cloud specific issue that can impact on availability of data.
		\paragraph{Resource accounting issues} PaaS systems enable third party applications to run on a shared platform (see also Section \ref{subsubsection:conf}). Resource accounting mechanisms are required in order to monitor and limit the applications utilization of resources. In \cite{Rodero_2012}, it was shown that both Java and .Net (which can both be used to implement a PaaS system) lacked of mechanisms for monitoring resources. This situation could have been exploited by malicious tenants to keep instantiating objects until the Cloud provider memory was exhausted.

	\subsubsection{Network level issues}
		As for the previous layer, even in this case we have identified only one Cloud specific issue located at the network level and capable of affecting Cloud availability.
		\paragraph{Network under-provisioning issues} A new form of DoS attack in Cloud scenarios that exploits network under-provisioning is described in \cite{Liu_2010}.

	\subsubsection{Multi-level issues}\label{subsubsection:Economic-vulnerabilities}
		In the next paragraph we present a class of attacks, also known as Economic Denial of Sustainability attacks, that have the potential to impact availability of services deployed on the Cloud. Since this class of attacks represents a methodology to strike a Cloud system, which can be implemented by exploiting several protocols located at more than one layer of our architecture, we have decided to present it in this parallel subsection and separated from the layer-oriented classification.
		\paragraph{Economic sustainability issues} this category represents a set of attacks aimed at causing financial burden for providers offering services through the Cloud\cite{Somani_2017} with the purpose of making the Cloud economically  unsustainable \cite{Ficco_2016}.

		An example of such attack is Fraudulent Resource Consumption (FRC). In this case, the adversary behaves as a normal user and requests to the victim's service deployed on the Cloud to perform some operations. However, differently form a flooding attack, the adversary does not seek to congest the  service provider resources; instead, he seeks to maintain a low profile of requests (i.e. produce a quantity of requests that will not be as overwhelming as is the case for flooding attacks) with the purpose of being able to produce them for a long period of time\cite{Xiao_2013}. As result, the adversary, exploits the pay as you go and auto-scaling models for billing to the service provider an unforeseen amount of resource utilization. The attacker's aim, is that, eventually, the service provider will face unexpected expenses which will lead to economic losses and therefore deprive the long-term economic availability of using the Cloud \cite{Xiao_2013}, which in turn may also result in a denial of service attack and make the targeted services unavailable on the Cloud \cite{Somani_2015}.

		When the resource consumed by an FRC attack is the electrical energy and power of the Cloud infrastructure, we refer to such an attack as Energy-related Denial of Service attack (e-DoS)\cite{Ficco_2017}. In this case the adversary's goal is to produce a limited amount of requests that will switch the victim's electronic facilities from low energy consumption states to high energy consumption states \cite{Ficco_2017}.

		As noted in \cite{Somani_2017}, a naive solution to these type of attacks, would be to disable the auto-scaling capabilities offered by the Cloud. However, with the lack of auto-scaling, the attack would directly result in a denial of service and would also nullify the advantages of the Cloud environment.

		Even if this category of attacks is not completely aimed at compromising availability of services, similarly to various works in literature (e.g. \cite{Xiao_2013}), we consider it as a problem of availability. The main reason behind this choice is related to the similarity that this attacks have with DoS attacks. Moreover, by making the Cloud economically disadvantageous, service provider may be pushed to remove their services from the Cloud and hence, in a Cloud perspective, factually render such service unavailable on it.

\section{Cloud-generic security issues}\label{sec:generic_cloud_issues}
In this section, we present a number of security issues that, despite being present in Cloud computing, are also common in many other paradigms. We consider them either because they are more dangerous in the Cloud environment than in other models, or because it is more difficult to solve them in the Cloud environment. Moreover, we do not focus on reporting vulnerabilities (since these are well-known), but we only report attacks we consider worth mentioning.

Three classes of security issues are treated: network-level issues, web-technologies issues, and multi-level issues. For each of them, the identified security issues are further categorized based on the CIA properties they affect. Hence, the flow of this section is reverted if compared to the previous one (Sec.~\ref{sec:specific_cloud_issues}): first, issues are grouped in relation to the architectural level they influence (i.e., network level, web-technologies, or multi-level), then, they are specialized in relation to the property they affect.

\subsection{Network level issues}
The security issues considered at this level are not specific of the Cloud but are common to every networked system because related to vulnerabilities at the transport, connection and data-link layers of the OSI-model. Nevertheless, there exist some specific characteristics of the Cloud that complicate the typical networking scenario and pose some challenges in finding proper solutions to such problems.
	\paragraph{Lack of a clear security perimeter} Most network protection techniques rely on the assumption that there exists a clear perimeter between an internal (trusted) and an external (untrusted) environment from which attacks are injected \cite{Shin_2012,Zissis_2012}. However, for a public Cloud this is not true since attacks can also be generated from within the Cloud and affect other clients of it. Therefore, security mechanisms can not be simply installed at the entry of the Cloud network (as it would also exhibit low scalability \cite{He_2016}) but instead it is needed to assess and evaluate where they have to be located and possibly permit tenants to deploy their own security mechanisms \cite{Shin_2012}.
	\paragraph{Network diversity} The network of a Cloud environment is characterized by a high degree of diversity and complexity. This is mainly caused by the mix of physical and virtual networks, by the different network configurations required from tenants\cite{Shin_2012} but also related to the different protection configurations required by all the various services deployed in the Cloud\cite{He_2017}. Therefore, such complexity might impact the effectiveness of security mechanisms. For instance, if we consider the case of listed-rules firewalls, the complexity of the Cloud network might lead to several errors in the definition of proper rules and therefore make it difficult to deploy such security mechanisms \cite{He_2014};
	\paragraph{Dynamic network topology} Since Virtual Machines (VMs) deployed in the Cloud can be migrated from one location to another, it is paramount to also enable the relocation of defense mechanisms (such as Network Intrusion Detection Systems, NIDSs) as the VMs move in such a way to guarantee that the traffic produced by a given VM is still controlled independently from its location \cite{Shin_2012,Wang_2015}.

	\subsubsection{Confidentiality and Integrity issues}
	Well-known examples of network-level attacks that can affect the confidentiality and integrity of every networked system are: packet sniffing, IP spoofing, ARP spoofing and Man In The Middle attacks (MITM) \cite{Coppolino_2017,Modi_2013,Kim_2014}. Among network level issues that can affect integrity and confidentiality, MITM attacks are much more critical and deserve special attention.
		\paragraph{Man-In-The-Middle attacks (MITM)} The entities involved in a MITM attack are two nodes seeking to communicate with each other and an adversary having access to the communication channel connecting the two endpoints. The adversary exploits its access to the communication channel for intercepting and modifying messages exchanged between the two victims, in such a way to make them believe that they are securely communicating with each other, while, in reality, they are actually exchanging messages only with the attacker who can properly modify and redirect messages to hide its presence\cite{Nicola_2016}.

		Since the Cloud heavily relies on networks and remote communications, it is clear that MITM attacks are paramount attacks in this scenario as can affect confidentiality, integrity and also availability of data.

	\subsubsection{Availability issues}\label{subsubsec:Network-availability}
		We are here going to review the main attacks at network level that can impair availability. As will be noted at the end of the following paragraphs, even if DoS attacks are common also to many other paradigms, their effects in a Cloud environment are actually different and more subtle if compared to other existing models.

		\paragraph{Distributed Denial of Service (DDoS) attacks} DDoS attacks are typically performed by means of remotely controlled devices (also know as bots or zombies) that are coordinated to simultaneously generate a multitude of requests that will eventually flood the victim's resources \cite{Zargar_2013,Osanaiye_2016,mido_2017_iSociety}. The employment of bots allows to: hide the original attacker, make it difficult to identify when an attack is being performed (i.e. it is difficult to understand that the different incoming requests are actually coordinated by the same source) and generate a huge quantity of traffic \cite{Zargar_2013,mido_2017_INSERT} which makes the attack much more disruptive than a simple SDoS \cite{Chapade_2013}. To have an insight of the potential power of such attack, it is sufficient to consider that the recent Mirai malware allowed to generate traffic peaks of more than 620 Gbps \cite{Kolias_2017,mido_2018}.

		Transport and network level protocols can be used to exhaust network resources \cite{Zargar_2013}. Typical examples of network-level DDoS attacks that can affect the Cloud are\cite{Osanaiye_2016,Miao_2015}:
		\begin{itemize}
			\item \textit{TCP SYN flooding attack}: in this case, an adversary takes advantage of the TCP three-way handshake mechanism, for creating various \quotes{half-opened connections} \cite{Hawawreh_2017}. These half-opened connections will be initially stored in a backlog queue but when its maximum size is reached, the server will deny all subsequent connections\cite{Hawawreh_2017} making its services unavailable for other clients;
			\item \textit{UDP flooding attack}: in this type of attack the connectionless and unreliable features of UDP are exploited to make the target system unreachable \cite{Osanaiye_2016}. An example of UDP flooding attack is the UDP storm attack \cite{Gupta_2017_DDoS}.
			\item \textit{ICMP flooding attacks}: an example of such attacks is the Ping flooding attack. In a Ping flooding attack the adversary seeks to congest the victims' network by sending massive ICMP echo requests packets to which the victim he has to reply back \cite{Chapade_2013}.

			In some cases, ICMP flooding may also be implemented in the form of a SMURF attack \cite{Wong_2014}.
		\end{itemize}

		The analysis performed in \cite{Miao_2015} demonstrates that the aforementioned attacks are a considerable threats for Cloud computing. Indeed, it is shown that the Cloud can be both the victim of those attacks but it could also be exploited to perform them (an example of malicious Cloud exploitation are botClouds\cite{Badis_2014}, i.e. botnets deployed in the Cloud environment).  Moreover, the authors of \cite{Yan_2016}, have identified an increase in the number of occurrences of such attacks in Cloud environments. According to \cite{Somani_2017} and \cite{Yan_2016}, the reason for that may be rooted in the intrinsic characteristics of the Cloud which, in a certain way, can support the success of DDoS attacks. For this reason, in \cite{Yan_2016}, the authors correlate DDoS attacks with the Cloud essential characteristics in order to discuss the reasons that make Cloud computing a juicy target (but even a source) of such attacks.

		The authors of \cite{Jensen_2009}, identified two types of flooding attacks effects which are specific of the Cloud:
		\begin{itemize}
			\item \textit{Direct DoS}: when a service deployed in the Cloud is targeted by a flooding attack, the Cloud provider will increase the victim's resources for allowing the victim to hold up against the attack. However, by doing so, the Cloud provider also supports the attack since enables the adversary to impair an entire service availability by just flooding one single node that is part of the service;
			\item \textit{Indirect DoS}: a flooding attack toward a service located in the Cloud, does not only affect the direct target of it, but it can also impair the availability of all the other services that are sharing the same machine with the victim. To demonstrate this, the authors of \cite{Somani_2015,Somani_2016} performed system analysis and simulations to show that different Cloud stakeholders are affected by Cloud DoS attacks even if they are not explicitly targeted as victims.
		\end{itemize}

\subsection{Web technologies issues}
SaaS services are typically delivered to end-users by means of web browsers \cite{Modi_2013}, while PaaS services are typically accessed through Web Services \cite{Jensen_2009}, \cite{AlMorsy_2011}. It follows from the use of these access technologies that web technology weaknesses are also inherited by the Cloud.
	\subsubsection{Confidentiality issues}\label{subsection:Confidentiality_Web}
		We report here some of the main confidentiality issues that are rooted at web-technology layer.
		\paragraph{Cross-site scripting (XSS) attacks} depending on the way it is performed, different specialization of this attack exist \cite{Gupta_2017}. However, the general idea behind an XSS attack is to exploit web-server vulnerabilities to inject JavaScript code within web-pages that are later going to be downloaded by victims. Once the web-page is accessed, the malicious JavaScript code therein contained will be executed on the client's browser\cite{Gupta_2017}, \cite{Shar_2012}. This threat can potentially allow to steal cookies \cite{Putthacharoen_2011} (and subsequently lead to session hijacking \cite{Dacosta_2012}), steal user's access credentials by placing forged input forms within targeted web sites (i.e. phishing attack) or set up keystroke based attacks which can allow the attacker to infer passwords or other credential\cite{Gupta_2017}. All the aforementioned attacks have the common consequence of allowing an unauthorized party to access victims data.

		\paragraph{Injection flaws} in these types of attacks, input sources are exploited to inject malicious input data that, once interpreted, can influence the execution of back-end operations and change their expected results \cite{SUBASHINI_2011}, \cite{Grobauer_2011}. This type of attack is particularly relevant in SaaS applications where data of different users are stored in common structures and where intrusion risks are amplified \cite{SUBASHINI_2011}. Various examples of injection attacks exist: SQL injection, command injection (also known as OS injection), XML injection and many others. As regards confidentiality, injection flaws can lead to unauthorized access of data and information disclosure \cite{OWASP_2017}.

		\paragraph{Man-in-the-Browser attack (MitB)} Man in the Browser attacks are similar to Man-In-The-Middle-Attacks. The difference among them is related to the level in which they operate to intercept data and perform the attack. While Man-In-The-Middle attacks are typically performed at network level, Man-in-the-Browser attacks are performed at application level \cite{Kim_2014}.
		By operating at the application level, these attacks have the capability of stealing confidential information that transits through browsers \cite{Morrow_2012}.

	\subsubsection{Integrity issues}
		The focus in the next paragraphs is on web-related attacks that can impair integrity of data and elaboration.
		\paragraph{Cross-site request forgery (CSRF) attack} when users have an active connection with a trusted web server and are simultaneously visiting another not trusted web page, an attacker owning the latter can exploit the session information contained in the users' browsers to forge unauthorized requests toward the trusted server as if they have been requested by the victim \cite{Shahriar_2010,Siddiqui_2011}. Since operations performed are not actually requested by the real owner of data, such an attack can affect the integrity of user's data contained in the Cloud.

		\paragraph{Hidden field manipulation and cookie poisoning} due to the stateless nature of HTTP, hidden fields and cookies are usually adopted to keep track of sessions. Hence, a modification of their content can allow to inject malicious data within web applications and consequently break the integrity of data stored in web servers  \cite{Livshits_2005}, \cite{SUBASHINI_2011}. In some cases, cookie poisoning attacks are also performed with the objective of obtaining access to unauthorized resources \cite{You_2012}.

		\paragraph{Injection flaws} these attacks (see also confidentiality paragraph in Subsection \ref{subsection:Confidentiality_Web}) can also affect the integrity of data and cause data corruptions\cite{OWASP_2017}.

		\paragraph{Man-in-the-Browser attack (MitB)} In Main-in-the-browser attacks (see also confidentiality paragraph in Section \ref{subsection:Confidentiality_Web}) a trojan horse is placed within victim's host (typically in its browser, but it is not always the case as some variations exist) which allows the intruder to intercept and modify network traffic \cite{Rauti_2014}. Once a valid session has started, the malware can tamper data sent from both the user and remote server in such a way to alter requests and responses as he prefers while potentially leaving both parties unaware of such situation \cite{Rauti_2014}.

		\paragraph{XML Signature Element Wrapping} this type of attack allows to break the integrity of SOAP messages when the XML Signature standard is naively used\cite{Kim_2014}. Indeed, starting from a signed SOAP message it is possible to wrap it within the body of a new message and let the web service believe that it has been generated by the legitimate user \cite{Jensen_2009}. In the Cloud context, Amazon EC2 has been in the past identified vulnerable to this attack \cite{Jensen_2009}.

		\paragraph{Metadata Spoofing attack} the WSDL standard is used to describe services in the SOAP model. A metadata spoofing attack aims at modifying a service WSDL in such a way that requests created according to the malicious WSDL result in elaborations different from those intended by the designers of the attacked web service \cite{Jensen_2009}.

	\subsubsection{Availability issues}
	The following paragraphs present some of the most important attacks on availability that take advantage of web technology vulnerabilities. We have chosen to not emphasize the difference among SDoS and DDoS attacks, however, we distinguish between the type of attacks: application-enabled DoS and flooding attacks. In relation to such attacks, it should be noted that the discussion we develop here is an extension of subsection \ref{subsubsec:Network-availability} where, in this case, the focus is on denial of service attacks that are performed through exploitation of application level protocols.

	\paragraph{Application-bug level DoS} rather than representing a specific type of attack, this entry represents a set of DoS attacks which are performed by exploiting protocol vulnerabilities, system weaknesses, mis-configurations or lack of updates\cite{Osanaiye_2016}. What distinguishes this type of attacks from from other types of DoS (such as flooding attacks) is that they can usually be easily solved with the installation of patches solving the identified vulnerabilities \cite{Beitollahi_2012}. Examples of this type of issues are:
	\begin{itemize}
		\item \textit{HTTP PRAGMA} and \textit{HTTP POST attacks}: in both of them, the adversary, takes advantage of specific HTTP requests in order to keep consuming and maintaining control of much more resources than actually needed \cite{Dantas_2014};
		\item \textit{Coercive parsing}: once received, SOAP messages have to be parsed by web services. However, in some cases such operation might become highly time-consuming and complex due to XML\cite{Jensen_2009_2}. Therefore, attacker may create deeply nested XML structures in such a way to exhaust computational resources \cite{Jensen_2009_2}. Hence, according to the definition of resource depletion attack given in \cite{Specht_2004,Shamsolmoali_2014}, we identify this type of attack as a DoS attack. In \cite{Rak_2015}, this type of attack has been used as a case study for performing Economic Denial of Sustainability attacks (see also Section \ref{subsubsection:Economic-vulnerabilities}) within Cloud environments;
		\item \textit{Chained encrypted keys}: WS-Security describes enhancement for providing integrity and confidentiality of SOAP messages by means of flexible mechanisms which can be negotiated by the parties involved in the communication. An attacker can take advantage of such flexibility to attack service provider availability by creating a SOAP message containing a nested sequence of encrypted keys which the web service need to recursively decrypt in order to find the final key and perform the elaboration\cite{Jensen_2009_2}. A similar attack which exploits the same vulnerability is nested encryption blocks \cite{Jensen_2009_2}.
	\end{itemize}

	\paragraph{Flooding attacks} Differently from the other category of DoS attacks, in this case the attacker does not exploit any vulnerability at application level but seek to exhaust the target resources by producing an overwhelming quantity of requests \cite{Osanaiye_2016}. Some examples of these attacks that can affect the Cloud are:
	\begin{itemize}
		\item \textit{HTTP flooding attack}: in this type of DoS attack, the adversary's goal is to generate large amounts of HTTP GET (or HTTP POST) requests that will eventually fill-up the victim's request queues\cite{Singh_2017,Choi_2014,Osanaiye_2016}. As the victim's resources have been occupied to satisfy requests from the attacker, this will lead to a rejection of all the subsequent requests performed by honest users;
		\item \textit{XML Oversize Payload attack}: in an oversize payload attack a saturation of available resources is achieved by submitting to service providers large queries\cite{Jensen_2009_2}. Since XML files are characterized by significant processing overhead  they might be exploited to saturate resources and compromise services availability \cite{Kim_2014,Jensen_2009_2}.
	\end{itemize}

	\paragraph{Injection flaws} (introduced in subsection \ref{subsection:Confidentiality_Web}): data loss or denial of access can be performed by means of injection flaws\cite{OWASP_2017} with the consequence of compromising data and service availability.

\subsection{Multi-level issues}

In this category we present one type of issue that is not specific of any layer of our reference architecture but that according to its specific implementation it might embrace more than one level. Moreover it is not even specific of the Cloud environment however it has the characteristic of being tailored to the environment it attacks, and therefore its implementation might in reality be a form of Cloud specific attack.

\subsubsection{Confidentiality issues}

\paragraph{Advanced Persistent Threats} two are the features that make Advanced Persistent Threats (APT) different from viruses and worms: 1) they are tailored for a specific target user and 2) the attacker is patient, i.e. willing to invest lots of resources and time for accomplishing his mission. After a first phase of information gathering aimed at identifying the victim's behaviors, network and defenses, the enemy develops a tailored attack with the ultimate objective of loading a malware into the victim's target machine in such a way to extract its information \cite{Sood_2013,Chen_2014}.

The Cloud could be affected by such types of attacks in two possible ways: it can either be exploited for performing the attack by silently transmitting information from the victim to the attacker, i.e. covert channels enabled by the Cloud are used \cite{Caviglione_2017}, or it could be directly attacked and used as platform for hosting malware\cite{Sood_2013} with the objective of stealing Cloud users' data for long period of times and without being noticed \cite{Xiao_2017}. The last case is particularly dangerous for a Cloud environment because once a user gets infected it can also compromise other services and users\cite{Sood_2013}.

\section{Conclusion and Future work}
\label{sec:conclusion}
In this paper, the security of IoT has been analyzed from a specific perspective: Cloud computing considered as a core component of the IoT architecture. The motivation behind this work resides on the evidence that, today, IoT devices strongly rely on the Cloud, where data analytics and intelligence reside. Therefore, addressing security of IoT devices and Cloud computing as different concerns is no longer enough to tackle security issues of the IoT, in its broader meaning.
As a result, we have provided an up-to-date and well-structured survey of the security issues of Cloud computing in the IoT era. The analysis has been based on a structured approach, distinguishing between Cloud-specific and Cloud-generic security issues, and classifying both classes from two angles: the affected Cloud architectural layer and the impacted CIA security property (i.e., confidentiality, integrity, availability). We believe that this classification is important to have a clear picture of where security issues occur and what their potential impact is. As a result, our analysis points out that, since there is no IoT without the Cloud, we cannot secure IoT without securing the Cloud. Thus, we consider this work as a first step toward the investigation of IoT security in its broader meaning.

This work can be extended in different ways. For instance, it could be useful to add a risk analysis, specifying the risk associated with each vulnerability.
Moreover, due to the broad nature of the topic covered in this paper, we have tried to keep its scope very well focused, considering only the fundamental and well-known CIA security properties. Nevertheless, it would be interesting to extend the analysis by taking into consideration other relevant security properties, such as authenticity and accountability.

Looking at Microservices as an architectural approach for creating cloud applications, where each application is designed and built as a set of services defined by business capabilities, the analysis could expand into this domain and the related programming languages \cite{GuidiLMM17}. Microservices and IoT, and related security challenges, have certainly lot in common with what described in his work, but certain peculiarities would deserve a separate discussion. Formal approaches and rigorous semantics have also not been considered in this work despite their importance for Cloud and distributed/concurrent systems in general \cite{YanCZM07,Mazzara:phd,DragoniM09}.


\bibliographystyle{abbrv}

\bibliography{myBibliography}

\end{document}